\documentclass[conference]{IEEEtran}
\IEEEoverridecommandlockouts
\usepackage{cite}
\usepackage{amsmath,amssymb,amsfonts}
\usepackage{algorithmic}
\usepackage{graphicx}
\usepackage{textcomp}
\usepackage{xcolor}
\usepackage{xspace}
\usepackage{xurl}
\def\BibTeX{{\rm B\kern-.05em{\sc i\kern-.025em b}\kern-.08em
    T\kern-.1667em\lower.7ex\hbox{E}\kern-.125emX}}
\begin{document}

\newcommand{\elasticai}{ElasticAI\xspace}
\newcommand{\creator}{ElasticAI-Creator\xspace}
\newcommand{\node}{Elastic Node\xspace}
\newcommand{\workflow}{ElasticAI-Workflow\xspace}

\title{\elasticai: Creating and Deploying Energy-Efficient Deep Learning Accelerator for Pervasive Computing
}

\author{\IEEEauthorblockN{1\textsuperscript{st} Chao Qian}
\IEEEauthorblockA{\textit{Intelligent Embedded Systems Lab} \\
\textit{University of Duisburg-Essen}\\
Duisburg, Germany \\
chao.qian@uni-due.de}
\and
\IEEEauthorblockN{2\textsuperscript{nd} Tianheng Ling}
\IEEEauthorblockA{\textit{Intelligent Embedded Systems Lab} \\
\textit{University of Duisburg-Essen}\\
Duisburg, Germany \\
tianheng.ling@uni-due.de}
\and
\IEEEauthorblockN{3\textsuperscript{rd} Gregor Schiele}
\IEEEauthorblockA{\textit{Intelligent Embedded Systems Lab} \\
\textit{University of Duisburg-Essen}\\
Duisburg, Germany \\
gregor.schiele@uni-due.de}
}
\maketitle
\begin{abstract}
Deploying Deep Learning (DL) on embedded end devices is a scorching trend in pervasive computing. Since most Microcontrollers on embedded devices have limited computing power, it is necessary to add a DL accelerator. Embedded Field Programmable Gate Arrays (FPGAs) are suitable for deploying DL accelerators for embedded devices, but developing an energy-efficient DL accelerator on an FPGA is not easy. Therefore, we propose the \workflow that aims to help DL developers to create and deploy DL models as hardware accelerators on embedded FPGAs. This workflow consists of two key components: the \creator and the \node. The former is a toolchain for automatically generating DL accelerators on FPGAs. The latter is a hardware platform for verifying the performance of the generated accelerators. With this combination, the performance of the accelerator can be sufficiently guaranteed. We will demonstrate the potential of our approach through a case study.
\end{abstract}

\begin{IEEEkeywords}
    Deep Learning, FPGA, Energy-Efficiency
\end{IEEEkeywords}

\section{Introduction and Related Work}
Introducing Deep Learning (DL) to embedded devices enables things to perceive and react to their surroundings with intelligence. Cloud/Edge-based DL approaches are now well-established and widely adopted, but they have shortcomings in accessibility, reliability, and privacy. Consequently, there is a tendency to deploy DL models on end devices, where Microcontrollers (MCUs) are often employed as the processor to deal with general tasks due to their power efficiency. However, their low-power feature results in the lack of resources and computational capacity required for the efficient execution of DL models. Sometimes, highly optimized models infer too slowly on such MCUs to reach real-time requirements.

Previous studies \cite{medus2019novel,roggen2022wearable} indicate that embedded Field Programmable Gate Arrays (FPGAs) are advantageous for accelerating algorithms to gain greater speed or power efficiency. Thus, we use an FPGA to execute highly optimized, application-specific DL accelerators instead of employing a more powerful MCU for faster DL operations. The low-power MCU is then always on and in charge of arranging onboard resources. At the same time, the FPGA can be activated to accelerate DL model inference when needed and will be deactivated during idle states to conserve energy. By leveraging features from the MCU and FPGA, the device can gain DL acceleration while meeting energy-efficiency requirements for the entire system.

To infer a DL model on an FPGA, a corresponding DL hardware accelerator should be implemented. Many approaches \cite{fahim2021hls4ml,xilinxXilinxVitisAI2022} use High-Level-Synthesis (HLS) to simplify the development of DL accelerators on FPGAs. 
However, it is known that HLS introduces significant overhead. Blott et al. \cite{blottFINNREndtoEndDeepLearning2018} report that HLS introduced 45\% resource overhead in their case. Thus, we prefer to create DL accelerators using Register Transfer Level (RTL) template-based approaches. Although designing such templates incurs additional development workload, it is worthwhile for low-energy devices since it saves more resources and energy than HLS approaches.

With software, estimating the power consumption of generated DL accelerators is feasible. However, noticing the unreliability of estimations, researchers also measure power consumption on real hardware \cite{roggen2022wearable, khabbazan2019design,chen2021eciton,burger2020embedded,burger2018demo}. Nevertheless, major hardware platforms can only provide their overall power consumption, which is insufficient to verify the power estimation of the accelerator, let alone to guide the optimization of the accelerator. Thus, we favor developing special hardware to acquire fine-grained power consumption measurements.

In this work, we propose the \workflow, which consists primarily of the \creator toolchain \cite{chao2022creator} and the \node \cite{burger2018demo} hardware platform. We plan to demonstrate the following:
\begin{itemize}
    \item a workflow (the \elasticai) to help DL developers to create and deploy energy-efficient DL accelerators,
    \item an interactive process of solving a certain DL task using the workflow to show the potential of our approach.
\end{itemize}

In the next section, we will introduce the details of our system. Then, a brief description of our demo will be provided. Finally, we will conclude with some thoughts and propose future work.

\section{System Description}
This section first discusses the challenges of developing energy-efficient DL accelerators on embedded FPGAs and then explains our approach to structuring the \workflow.

\subsection{Challenges}
We have been researching emerging DL implementations for embedded devices that own low-power MCU and FPGA for several years. Reflecting on our experiences, we notice two critical challenges to achieving energy efficiency.

\paragraph*{1) Solid FPGA expertise is required} 
To fit embedded FPGAs, DL developers need to aggressively optimize the DL model to simplify computation and reduce memory footprint while preserving the model's ability to give valid results. Then, to execute a particular (optimized) model on FPGAs, a corresponding digital circuit (accelerator) must be implemented, where optimizations for energy efficiency at the hardware level are required. For example, the constrained resources make it impossible to naively instantiate all components of the DL model on an embedded FPGA to run all components in parallel. Instead, reusing components over time to save resources is often applied for such FPGAs. Therefore, the developers must have  sufficient FPGA engineering expertise to create implementations that maximize the benefits of adding an FPGA by balancing resource utilization and energy efficiency. However, it is non-trivial and time-consuming for DL developers to master this knowledge.

\paragraph*{2) Reliable and fine-grained power consumption measurements are required} 
Our previous work \cite{qianenhancing} indicates that before running an accelerator on an FPGA, it is possible to get a performance estimation by simulating the accelerator on the circuit level and logging all switching activities using Vivado\footnote{Xilinx Vivado: https://www.xilinx.com/products/design-tools/vivado.html}. However, the estimated power consumption can be inaccurate even with adequate simulation data. In addition, such simulations perform intensive computation and may take a long time. A task that only takes a few seconds to execute on real hardware may take hours in such a simulation. Therefore, we prefer to perform real-world measurements on hardware to ensure reliability. To provide a good granularity of power consumption measurements, similar to the ones estimated by Vivado, special hardware with the capability of separately measuring the power consumption for each function region is required.

\subsection{\creator}
\label{sec:creator}
To help DL developers cope with the lack of FPGA engineering knowledge, we propose the \creator, which extends PyTorch with model components that are translatable to RTL components for FPGAs. Once the developers have designed a DL model with the supported components, this model can easily be translated into a hardware accelerator. 
The trained and optimized model can be translated to a hardware accelerator in the RTL representation by simply pressing a button. Since the translation process can be taken care of by the \creator, the developers do not need to understand how FPGAs work.

\subsection{\node}
\begin{figure}[!htb]
	\centering
	\begin{minipage}{.33\columnwidth}
		\centering
		\includegraphics[width=\textwidth]{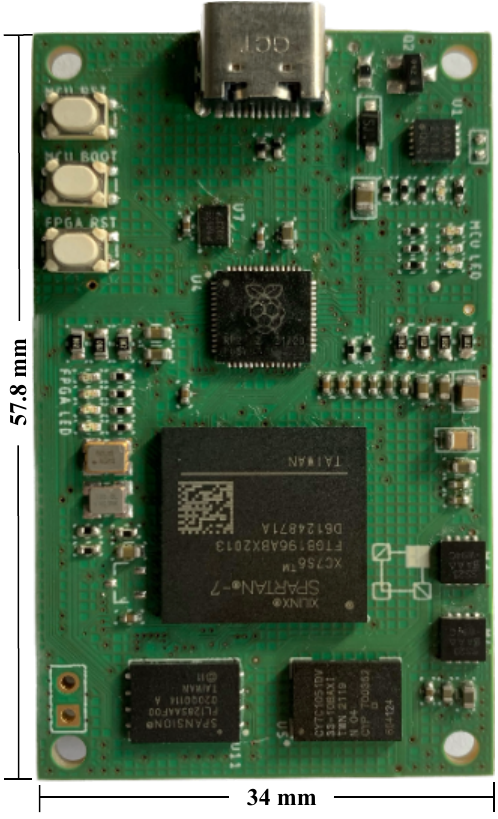}
		\caption{Elastic Node V5}
		\label{fig:en_v5_pcb}
	\end{minipage}
        \hspace{.03\columnwidth}
 	\begin{minipage}{.55\columnwidth}
		\centering
		\includegraphics[width=\textwidth]{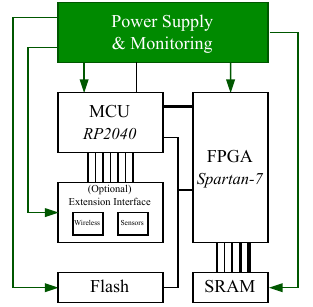}
		\caption{Elastic Node V5 System Diagram}
		\label{fig:en_v5_diagram}
	\end{minipage}%
\end{figure}

The \node is a customized hardware platform that can conveniently verify the energy efficiency of DL accelerators in a real-world environment. In 2018 we demonstrated the second version of the \node in PerCom \cite{burger2018demo} and showed live performance measurements when a DL model was executed on it. Since then, the \node has been upgraded to introduce new features to keep it advanced and fit more application cases. We now have the fifth version, which measures 57.8 $\times$ 34 mm, as shown in Figure \ref{fig:en_v5_pcb}. 

Figure \ref{fig:en_v5_diagram} demonstrates the system diagram of the hardware. The RP2040 MCU will be activated most of the time to control the sensors and the FPGA. The Spartan-7 FPGA belongs to the Xilinx product series with the lowest density. It is much more potent than our MCU but can still be battery-powered. 

A 320mAh rechargeable Lipo battery is connected to the Power Supply \& Monitoring block, enabling the device's portability. Inside this block, two energy meters PAC1934 are placed to monitor eight independent channels from all function regions on the \node. Each channel will contribute independent measurements to characterize the energy cost of each function region, guiding us to optimize the energy efficiency of DL accelerators. Last but not least, on the backside of the board, an extension interface can connect a wireless module to reach the Cloud and nearby nodes.

\subsection{\workflow}
\begin{figure}[!htb]
    \centering
    \includegraphics[width=0.85\columnwidth]{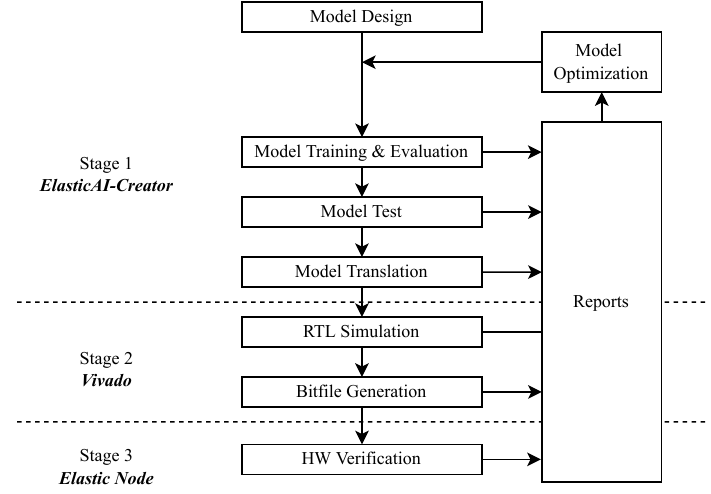}
    \caption{Simplified \workflow}
    \label{fig:elasticai_workflow}
\end{figure}

Combining the \creator and \node, we define the \workflow (see Figure \ref{fig:elasticai_workflow}) to develop energy-efficient DL accelerators on FPGAs. Currently, the \workflow involves three stages with a feedback loop that could start from reports on multiple levels. According to the information provided in the reports, DL developers can intervene with these stages to further optimize the DL model.

\textbf{In the first stage}, the DL developer can design a DL model with components supported by the \creator. After model training/evaluation and testing under the PyTorch Framework, this model can be automatically translated to an RTL representation by the \creator. Model optimization (such as hyper-parameters tweaking, layer replacement, and model quantization) can start with the information given in reports. The optimization loop will not terminate until the developers are satisfied with the reports.

\textbf{In the second stage}, the RTL representation of the generated accelerator is passed to the IDE of the FPGA vendor (such as Vivado from Xilinx) to synthesize and generate a bitfile that can configure the FPGA. With the aid of the IDE, more accurate resource utilization and power consumption estimations can be reported. Moreover, with RTL simulation, the inference timing can be extracted. Further, we can calculate the energy consumption/efficiency of the accelerator.

\textbf{In the last stage}, the accelerator can be executed on the hardware. During execution, the monitoring sub-system measures the power consumption in real time. The workflow is finished only when the hardware verification fulfills the requirement of the application. Furthermore, as the accelerator can infer very fast on the FPGA, it is feasible to measure its performance on the whole test set.

Please note that merging the first two stages is possible. We decouple these two stages to support FPGAs from other vendors.

\section{Demo Information}
To demonstrate how our \workflow works,  we will use a laptop and several Elastic Nodes to present it.
On the laptop, we will initially use the \creator in a Jupyter notebook to tune (optimize) a model design and perform model training with evaluation and testing for a particular DL assignment (e.g., time series analysis). 
Then, the \creator will translate the optimized model into an accelerator, which Vivado will use to generate the bitfile for FPGAs. The performance of the accelerator will be measured by loading the bitfile to the \node. Reports from various stages will be visualized and analyzed. In particular, measurements on the \node will be shown in real-time on the laptop. 

In addition, generating the bitfile takes several minutes (depending on the RTL design). While waiting, the audience can suggest modifications to each stage of the workflow. The performance of the newly generated accelerator will then be measured and compared to that of the previously generated accelerators. Potential reasons for performance changes can also be discussed.

\section{Experimental Results}
Table \ref{tab:exp_results} presents preliminary measurements from our previous work \cite{qianenhancing} of an accelerator generated by a Long Short-Term Memory model for predicting traffic flow. The estimation is slightly higher than real measurements. 
\begin{table}[!htb]
\centering
\caption{Experimental Results on XC7S15 @ 100MHz \cite{qianenhancing}}
\begin{tabular}{l|c|c}
                            & From Estimation    & From \node   \\ \hline
Power (mW)                  &  70                & 71           \\ 
Time per inference($\mu$s)  & 53.32              & 57.25        \\ 
Energy efficiency (GOP/J)   & 5.04               & 5.33         \\ 
\end{tabular}
\label{tab:exp_results} 
\end{table}

\section{Conclusion and Outlook}
Using \workflow, DL developers without FPGA expertise can create and deploy energy-efficient DL accelerators on embedded pervasive devices. Moreover, the accelerator's performance is promised due to the inclusion of verification on \node. Consequently, DL developers can concentrate on enabling more DL models to be deployed on embedded devices while addressing a broader range of target applications. In the near future, we intend to develop further model components to support more prevalent DL models, generating more DL accelerators. Moreover, we want to support FPGAs from other vendors. 

\bibliographystyle{IEEEtran}
\bibliography{reference}

\begin{thebibliography}{10}
\providecommand{\url}[1]{#1}
\csname url@samestyle\endcsname
\providecommand{\newblock}{\relax}
\providecommand{\bibinfo}[2]{#2}
\providecommand{\BIBentrySTDinterwordspacing}{\spaceskip=0pt\relax}
\providecommand{\BIBentryALTinterwordstretchfactor}{4}
\providecommand{\BIBentryALTinterwordspacing}{\spaceskip=\fontdimen2\font plus
\BIBentryALTinterwordstretchfactor\fontdimen3\font minus
  \fontdimen4\font\relax}
\providecommand{\BIBforeignlanguage}[2]{{%
\expandafter\ifx\csname l@#1\endcsname\relax
\typeout{** WARNING: IEEEtran.bst: No hyphenation pattern has been}%
\typeout{** loaded for the language `#1'. Using the pattern for}%
\typeout{** the default language instead.}%
\else
\language=\csname l@#1\endcsname
\fi
#2}}
\providecommand{\BIBdecl}{\relax}
\BIBdecl

\bibitem{medus2019novel}
L.~D. Medus, T.~Iakymchuk, J.~V. Frances-Villora, M.~Bataller-Mompe{\'a}n, and
  A.~Rosado-Mu{\~n}oz, ``A novel systolic parallel hardware architecture for
  the {FPGA} acceleration of {Feedforward Neural Networks},'' \emph{IEEE
  Access}, vol.~7, pp. 76\,084--76\,103, 2019.

\bibitem{roggen2022wearable}
D.~Roggen, R.~Cobden, A.~Pouryazdan, and M.~Zeeshan, ``Wearable {FPGA} platform
  for accelerated {DSP} and {AI} applications,'' in \emph{2022 IEEE
  International Conference on Pervasive Computing and Communications Workshops
  and other Affiliated Events (PerCom Workshops)}.\hskip 1em plus 0.5em minus
  0.4em\relax IEEE, 2022, pp. 66--69.

\bibitem{fahim2021hls4ml}
F.~Fahim, B.~Hawks, C.~Herwig, J.~Hirschauer, S.~Jindariani, N.~Tran, L.~P.
  Carloni, G.~Di~Guglielmo, P.~Harris, J.~Krupa \emph{et~al.}, ``hls4ml: An
  open-source codesign workflow to empower scientific low-power {Machine
  Learning} devices,'' 2021.

\bibitem{xilinxXilinxVitisAI2022}
Xilinx, ``Xilinx/{{Vitis-AI}},'' https://github.com/Xilinx/Vitis-AI, Sep. 2022.

\bibitem{blottFINNREndtoEndDeepLearning2018}
M.~Blott, T.~B. Preu{\ss}er, N.~J. Fraser, G.~Gambardella, K.~O'brien,
  Y.~Umuroglu, M.~Leeser, and K.~Vissers, ``{FINN-R}: An end-to-end {Deep
  Learning} framework for fast exploration of quantized {Neural Networks},''
  \emph{ACM Transactions on Reconfigurable Technology and Systems}, vol.~11,
  no.~3, pp. 16:1--16:23, 2018.

\bibitem{khabbazan2019design}
B.~Khabbazan and S.~Mirzakuchaki, ``Design and implementation of a low-power,
  embedded {CNN} accelerator on a low-end {FPGA},'' in \emph{2019 22nd
  Euromicro Conference on Digital System Design (DSD)}.\hskip 1em plus 0.5em
  minus 0.4em\relax IEEE, 2019, pp. 647--650.

\bibitem{chen2021eciton}
J.~Chen, S.~Hong, W.~He, J.~Moon, and S.-W. Jun, ``Eciton: Very low-power {LSTM
  Neural Network }accelerator for predictive maintenance at the edge,'' in
  \emph{2021 31st International Conference on Field-Programmable Logic and
  Applications (FPL)}.\hskip 1em plus 0.5em minus 0.4em\relax IEEE, 2021, pp.
  1--8.

\bibitem{burger2020embedded}
A.~Burger, C.~Qian, G.~Schiele, and D.~Helms, ``An embedded {CNN}
  implementation for on-device {ECG} analysis,'' in \emph{2020 IEEE
  International Conference on Pervasive Computing and Communications Workshops
  (PerCom Workshops)}.\hskip 1em plus 0.5em minus 0.4em\relax IEEE, 2020, pp.
  1--6.

\bibitem{burger2018demo}
A.~Burger and G.~Schiele, ``Demo abstract: {Deep Learning} on an {Elastic Node}
  for the {Internet of Things},'' in \emph{2018 IEEE International Conference
  on Pervasive Computing and Communications Workshops (PerCom
  Workshops)}.\hskip 1em plus 0.5em minus 0.4em\relax IEEE, 2018, pp. 424--426.

\bibitem{chao2022creator}
C.~Qian, L.~Einhaus, and G.~Schiele, ``{Elasticai-Creator}: Optimizing {Neural
  Networks} for time-series-analysis for on-device {Machine Learning} in {IoT}
  systems,'' in \emph{Proceedings of the 20th ACM Conference on Embedded
  Networked Sensor Systems}, 2022, pp. 941--946.

\bibitem{qianenhancing}
C.~Qian, T.~Ling, and G.~Schiele, ``Enhancing energy-efficiency by solving the
  throughput bottleneck of {LSTM} cells for embedded {FPGAs},'' in \emph{Joint
  European Conference on Machine Learning and Knowledge Discovery in
  Databases}.\hskip 1em plus 0.5em minus 0.4em\relax Springer, 2022, pp.
  594--605.

\end{thebibliography}
\end{document}